\documentclass[aps,prl,preprint,groupedaddress]{revtex4-2}
\usepackage{hyperref}
\usepackage{graphicx}

\begin{document}

\title{Platform and Framework for Time-Resolved Nanoscale Thermal Transport Measurements in STEM}

\author{Mairi McCauley\textsuperscript{1,2}}

\author{Joel Martis\textsuperscript{3}}
\author{Ondrej L. Krivanek\textsuperscript{3}}
\author{Ben Plotkin-Swing\textsuperscript{3}}
\author{Andreas Mittelberger\textsuperscript{3}}

\author{Tolga Wagner\textsuperscript{1,2}}
\author{H\"useyin \c{C}elik\textsuperscript{1,4}}
\author{Grigory Kornilov\textsuperscript{1}}
\author{Meng Zhao\textsuperscript{1,2}}

\author{Matthias Meffert\textsuperscript{5}}
\author{Luca Piazza\textsuperscript{5}}

\author{Tracy C. Lovejoy\textsuperscript{3}}

\author{Guillaume Radtke\textsuperscript{6}}

\author{Christoph T. Koch\textsuperscript{1,2}}

\author{Benedikt Haas\textsuperscript{1,2,*}}

\affiliation{\textsuperscript{1}Department of Physics, Humboldt-Universit\"at zu Berlin, Berlin, Germany}
\affiliation{\textsuperscript{2}Center for the Science of Materials Berlin, Humboldt-Universit\"at zu Berlin, Berlin, Germany}
\affiliation{\textsuperscript{3}Bruker AXS LLC, Kirkland, WA, USA}
\affiliation{\textsuperscript{4}Institute for Physics and Astronomy, Technische Universit\"at Berlin, Berlin, Germany}
\affiliation{\textsuperscript{5}DECTRIS AG, Baden-Daettwil, Switzerland}
\affiliation{\textsuperscript{6}Sorbonne Université, CNRS UMR 7590, MNHN, IMPMC, Paris, France}
\affiliation{\textsuperscript{*}Corresponding author. E-mail: benedikt.haas@hu-berlin.de }

\date{\today}

\begin{abstract}

Understanding heat transport at the nanometer scale is critical for semiconductor devices, quantum materials, and thermal management of nanostructures, yet direct local measurements of thermal conductivity and heat capacity remain scarce. We developed a laser-excitation system integrated into a scanning transmission electron microscope (STEM) for nanoscale thermal transport measurements using ultra-high-resolution electron energy-loss spectroscopy (EELS). A fiber-coupled laser is introduced via a modified aperture mechanism, enabling flexible holder geometries and large tilt angles without optical elements in the polepiece gap.
Synchronization of pulsed laser excitation with an externally gated direct electron detector provides temporal resolution about 50 ns at \textless10 meV energy resolution. Local temperatures are determined via the principle of detailed balance, and thermal transport parameters are extracted by fitting a forward-time central-space heat diffusion model including radiative losses.
For amorphous carbon films, we obtain a thermal conductivity of 1.24 \(\frac{W}{m\cdot K}\) and a heat capacity of 821 \(\frac{J}{kg\cdot K}\), consistent with literature. This framework enables time-resolved nanoscale measurements of thermal transport in materials and devices.

\end{abstract}

\maketitle

\subsection{Introduction}

Quantitative measurements of thermal transport properties at the nanoscale remain experimentally challenging. Thermal conductivity and heat capacity govern heat dissipation in nanoscale systems, yet direct measurements of these parameters with nanometer spatial resolution and sub-microsecond temporal resolution are scarce. Such measurements are advantageous to the design of efficient heat dissipation pathways in nanoscale devices e.g. via grain boundary engineering \cite{Kim2015_dislocation} and low dimensional systems \cite{ElSachat2021_lowD_thermal_transport}.

Scanning transmission electron microscopy (STEM) combined with ultra-high energy resolution electron energy-loss spectroscopy (EELS) now enables vibrational spectroscopy and nanoscale thermometry with energy resolutions below 10 meV \cite{Krivanek2014_vibEELS,Haas2024_Perspective,Hage2020_Superstem,Yan2021_Irvine,Qi2021_Gao,Xu2023_WuZhao,Hoglund2022_Oakridge}. Using the principle of detailed balance, local temperatures can be determined from the ratio of energy-loss and energy-gain phonon excitations \cite{Lagos2018,Idrobo2018}. Recent work demonstrated the measurement of interfacial thermal resistance using nanoscale thermometry and resistive heating to establish temperature gradients \cite{Liu2025_Gao_heating}. However, resistive heating approaches require complex microfabrication and restrict sample geometries.

Laser-based excitation provides an alternative route to induce localized heating in the electron microscope. A recent implementation employed a parabolic mirror inside the objective lens polepiece gap to focus laser light onto the sample and enabled nanosecond thermometry \cite{Castioni2025_OrsayNL}. While this approach provides high numerical aperture optical excitation, the integration of optical components in the polepiece gap restricts available space for sample holders and limits accessible tilt angles, reducing compatibility with in-situ biasing, cryogenic, and tomography experiments.

Here, we present a fiber-coupled laser injection system integrated into a Nion HERMES STEM via a modified aperture mechanism. The system introduces laser light into the microscope column without optical components inside the objective lens polepiece gap, maintaining full compatibility with a wide range of sample holders and large tilt angles. Laser excitation is synchronized with an externally gated acquisition mode of a Dectris ELA direct electron detector, providing temporal resolution about 50 ns while retaining high electron flux capability and \textless10 meV energy resolution.

Local specimen temperatures are determined using the principle of detailed balance applied to vibrational EELS and electron energy-gain spectra. To extract thermal transport parameters, we combine time-resolved temperature measurements with a forward-time central-space (FTCS) numerical heat diffusion model that includes radiative losses. This approach enables simultaneous determination of thermal conductivity and heat capacity from experimental data.

We demonstrate the framework on amorphous carbon thin films, obtaining thermal conductivity and heat capacity values consistent with literature. The presented instrumentation and analysis methodology provide a general platform for nanoscale thermal transport measurements in the electron microscope and are compatible with in-situ biasing, optical excitation, and tilt-series experiments.

\subsection{Materials and Methods}

\begin{figure}[htbp]
    \centering
    \includegraphics[width=0.95\linewidth]{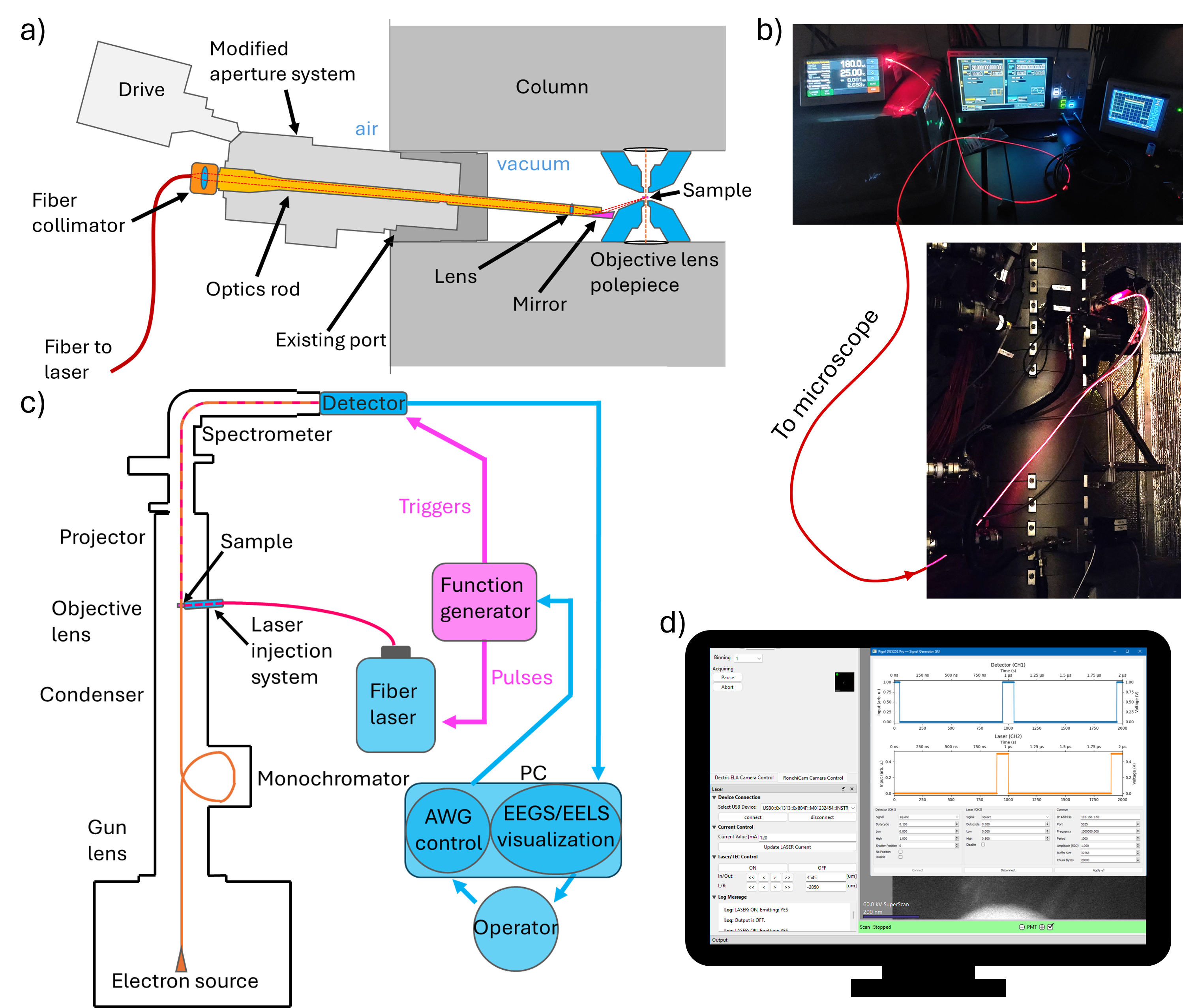}
    \caption{Experimental setup. a) Sketch of the laser injection system with a modified aperture mechanism attached to a port of the objective lens. b) Photograph of the laser controller, signal generator and oscilloscope setup (left) and of the microscope column with the laser fiber leading to the light insertion mechanism (right). c) Schematic of the microscope and time-resolved triggering system where a function generator is used to send synchronized time-delayed pulses to the laser and triggers to the detector. d) Screenshot of the laser plugin (motor and laser power control) integrated into the microscope control software (Nion Swift) and the homemade graphical user interface for the arbitrary wave generator (AWG).} 
    \label{fig:Setup}
\end{figure}

The laser injection system was integrated into a Nion HERMES scanning transmission electron microscope with a Dectris ELA direct detector attached to its IRIS spectrometer.
The basic idea of the setup is to use a modified aperture mechanism attached to an existing port of the objective lens. To this end, a hollow rod is inserted instead of the aperture holder which allows for the light to reach the sample (see Fig. \ref{fig:Setup}a). 
The output from a pigtailed laser diode is coupled via a single-mode fiber to a fiber collimator mounted at the exterior of the aperture mechanism, producing a parallel light beam with a diameter of approximately 3 mm that propagates along the optics rod. Near the sample, the rod terminates in a lens located about 30 mm upstream of the specimen which focuses the beam onto the sample. A flat mirror then redirects the focused beam onto the sample at an incidence angle of approximately 20$^\circ$ relative to the sample plane. A sapphire window at the back acts as the internal vacuum seal of the rod to the ultra-high-vacuum column and two o-rings, with differential pumping in between, seal the rod to the mechanism. This modified aperture mechanism allows for the optic rod to be translated with nm-level precision. The resulting horizontal displacement of the illuminated spot on the sample enables the laser illumination to be positioned relative to the location where the electron beam traverses the sample (e.g. to make them coincide). 

A photo of the outside of the microscope column with the optical fiber attached to the light injection system is shown in Fig. \ref{fig:Setup}b. The combined losses from the coupler, fiber and rod (collimator, lens, mirror) lead to about 70\% of the laser output power impinging on the sample. Unless otherwise stated, laser power refers to the laser output power not including these losses. For the shown experiments a red (637 nm wavelength) ThorLabs laser diode with peak power of 80 mW was controlled by a ThorLabs CLD1010LP laser controller. However, it is also possible to switch to other wavelengths within a few minutes by exchanging the laser diode.

Continuous wave laser power was controlled using a plugin implemented within the microscope control software (Nion Swift \cite{Meyer2019_NionSwift}) which is also used to control the position of the laser spot on the sample via aperture motors moving the optics rod inside the mechanism. Synchronized laser excitation and electron detection was achieved using a signal generator (Rigol DG5252 Pro) to power the laser and trigger the externally gated mode of the Dectris ELA detector with given frequency, duty cycle and phase shift. A sketch of the setup is shown in Fig. \ref{fig:Setup}c. The time resolution is limited by the detector response time to the external gate to about 50 ns. This value was determined by deflecting the electron beam using a pre-characterized \textit{in-situ} biased capacitor \cite{Wagner2019_capacitor} and is in agreement with data for the corresponding x-ray detector, Dectris Eiger \cite{Donath2023_Eiger2}. The detector integrates the impinging electrons during the gating signal for a given number of trigger pulses and produces a single frame. To maximize the signal-to-noise ratio without compromising the energy resolution, multiple frames were acquired, aligned and integrated to ensure sufficient exposure time without zero-loss peak broadening due to drift.

A silicon photodetector (ThorLabs DET025AFC) was used to characterize the laser pulses. As the damage threshold of this detector is 18 mW, the output power of the laser was reduced for the measurements and the data scaled up to the power values, determined by a thermal powermeter (ThorLabs PM160T), used in the experiment.

The principle of detailed balance (PDB) provides a parameter-free method of determining local temperature of a specimen within the microscope using the relative intensities in EELS and electron energy-gain spectra (EEGS) \cite{Idrobo2018,Lagos2018}. Photon absorption of the sample increases its temperature and the corresponding change in phonon population is related to this temperature via the Bose-Einstein distribution. 
The ratio of the intensity of electron energy-loss I(\textbf{q},E) (excitation of phonon) and energy-gain I(-\textbf{q},-E) (annihilation of phonon) peaks is related to the temperature by
\begin{equation}
	\frac{I(\mathbf{q},E)}{I(-\mathbf{q},-E)} = \text{exp}\left(\frac{E}{k_BT}\right),
\end{equation}
where $k_B$ is the Boltzmann constant and $E$ is the (phonon) energy. Linear fit of the ratio for different energies yields the temperature. Here we assume that the phonon population follows the Bose-Einstein statistics, i.e. the system has thermalized, which is well fulfilled at the given temporal resolution. Deviations from this would show up in a non-linear trend in the loss-gain ratio. Strictly speaking, the PDB does not hold for a temperature gradient (out of equilibrium state). However, up to large temperature gradients of typically thousands of $\frac{K}{\mu m}$, orders of magnitude larger than utilized here, the PDB still holds \cite{Liu2025_Gao_heating}.

To accurately determine the ratio between the loss and gain peak intensities, background subtraction of the zero-loss peak (ZLP) was carried out. Since many effects contribute to the ZLP intensity and lineshape, it was approximated using a Pearson VII function and fitted to four energy windows as shown by Levin et al. \cite{Levin2019_Pearson,Venkatraman2019_Pearson}. When a phonon peak was present in both the loss and gain spectra, the fit windows were selected on either side of the peak. More details about the analysis can be found in the results section below and in Fig. \ref{fig:Temp_Increase}. 
Where possible, an aloof spectrum was acquired and used to fit the ZLP. The accuracy of the background subtraction could be clearly observed when plotting the ratio of the gain/loss intensities as a poor fit resulted in large variations in values and the absence of a linear trend. 

Two different amorphous carbon films were used as samples. Continuous wave (CW) experiments were performed on a Quantifoil R1.2/1.3 film whose thickness was determined by the t/\(\lambda\) method \cite{Egerton2011_EELSbook} to be 16 nm. Time-resolved measurements were carried out on a continuous amorphous carbon film (S160-3, Plano GmbH) with a measured thickness of 8 nm.

A homemade Python script was used to simulate heat conduction of the continuous amorphous carbon sample using the actual geometry. We use the FTCS method to solve a 2D heat equation, similar to Castioni et al. \cite{Castioni2025_OrsayNL}. However, in addition to heat conduction we also consider radiative losses (as grey-body radiation) which lead to improved agreement between experiments and modeling. Due to the ultra-high vacuum in the microscope column, convection can be excluded. 
The central equation of the model contains a conduction term, laser heating term and radiation term (in this order) as
\begin{equation} \label{eq:2Dheat}
    \frac{\partial T(x,y,t)}{\partial t}=\frac{k}{\rho C_p}\nabla_{xy}^2T\ +\ \frac{{P_tAq}_{pulse}\left(x,y,t\right)}{\rho C_pd}-\frac{2R\varepsilon \sigma\left(T^4-T_{env}^4\right)}{\rho C_pd}
\end{equation}
where $k$ is the thermal conductivity, $d$ is the thickness of the sample, $P_t$ is the fraction of power transmitted to the sample, $A$ is the fraction of power absorbed by the sample (calculated from optical path length and absorption), $\rho$ is the mass density, $C_p$ is the heat capacity, $\varepsilon$ is the emission coefficient (fixed value for the whole spectral emission range; grey-body), the factor $2$ represents the two surfaces of the thin film (top and bottom), $R$ is a factor to account for its transparency as explained below and $\sigma$ is the Stefan-Boltzmann constant.
Dirichlet boundary conditions for $T_{env}$ = 300 K were used in the simulation to mimic the copper grid bars surrounding the 60x60 \(\mu\)m\textsuperscript{2} carbon film, acting as heat sinks. A 2D Gaussian distribution of the laser intensity on the sample was assumed and $q_{pulse}(x,y,t)$ spatially applies the laser power according to this distribution at each timestep.
We are determining the thermal conductivity and heat capacity (together with laser power losses, $P_t$, which also includes reflection) by optimizing the least square fitting of the model parameters to the experiment. For this approach we use the optical absorption at the laser frequency of approximately 14.6 \(\mu\)m\textsuperscript{-1}, the emission coefficient for the heat radiation of 0.8 and the mass density, 800 kgm\textsuperscript{-1}, for an amorphous carbon thin film \cite{Arakawa1977_aC_abs}. As the film thickness is significantly smaller than the optical absorption length of the material the assumption of two opaque surfaces (top and bottom) radiating independently is flawed. We therefore use an additional fit parameter in the radiative term, $R$, to account for this which is multiplied with the area. For an optically opaque thin film the value would be one (both surfaces radiating independently) but for transparent surfaces the factor is reduced (down to zero for a fully transparent surface). We obtain values of roughly 0.3 for the 8 nm continuous carbon film and about 0.45 for the 16 nm thick Quantifoil film.

\subsection{Results}

\begin{figure}[htbp]
    \centering
    \includegraphics[width=\linewidth]{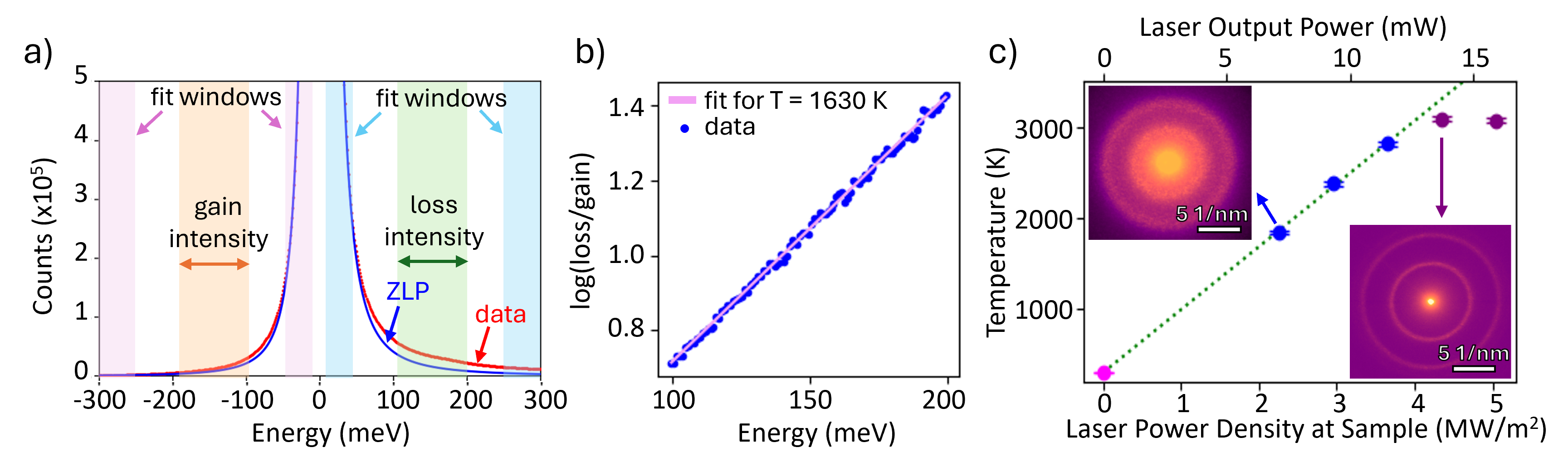} 
    \caption{Temperature analysis from EEGS/EELS measurements. a) EEGS/EELS spectrum with fitted ZLP and fit window indications. b) Plot of the natural logarithm of the loss/gain ratio against energy across the gain/loss intensity window indicated in a, showing a linear trend and corresponding fit of T = 1630 K. c) Measured temperature of a carbon film with continuous wave laser excitation of increasing optical power, where the pink data point is the assumed temperature with no laser stimulation (room temperature). The temperature increases linearly with increasing power (blue data points) from no optical power (pink) but deviates at high laser powers (purple) as the carbon structure is modified (diffraction patterns shown as insets).}
    \label{fig:Temp_Increase}
\end{figure}

The temperature of an amorphous carbon film of thickness 16 nm (see Materials and Methods) was measured for different CW laser powers at a single sample position close to the center of the laser spot. The specimen temperature was determined using PDB (see Materials and Methods) applied to the EEGS/EELS spectra from this position and is presented in Fig. \ref{fig:Temp_Increase}a-c. With laser output powers on the order of only 10 mW, the specimen was locally heated to above 2000 K. The film was observed to linearly heat up to ~3100 K starting from room temperature with increasing laser output power. Above this temperature, the linear trend no longer holds and the increase in sample temperature with increased laser power is reduced as seen in Fig. \ref{fig:Temp_Increase}c. We could observe that the diffraction pattern of carbon film heated above 3000 K, close to its evaporation temperature (around 3600 K), changes from a diffuse amorphous pattern to more defined rings associated with polycrystallinity as shown in Fig. \ref{fig:Temp_Increase}c. The fine structure of the carbon K edge in core-loss EELS shows a change from amorphous towards graphitic bonding configuration. As the thermal conductivity of graphite is significantly higher than amorphous carbon (in-plane around 300 compared to roughly 1 \(\frac{W}{m\cdot K}\)), even a small amount of graphitization should increase the thermal conductivity and, in a steady state, result in reduced heating localized to the laser spot. At even higher temperatures the carbon film was observed to evaporate instantaneously forming a hole at the center of the laser spot, which we will show and discuss below. This is in agreement with the evaporation temperature of carbon. These results give confidence in the ability of PDB to measure elevated temperatures at an absolute scale.

\begin{figure}[htbp]
    \centering
    \includegraphics[width=\linewidth]{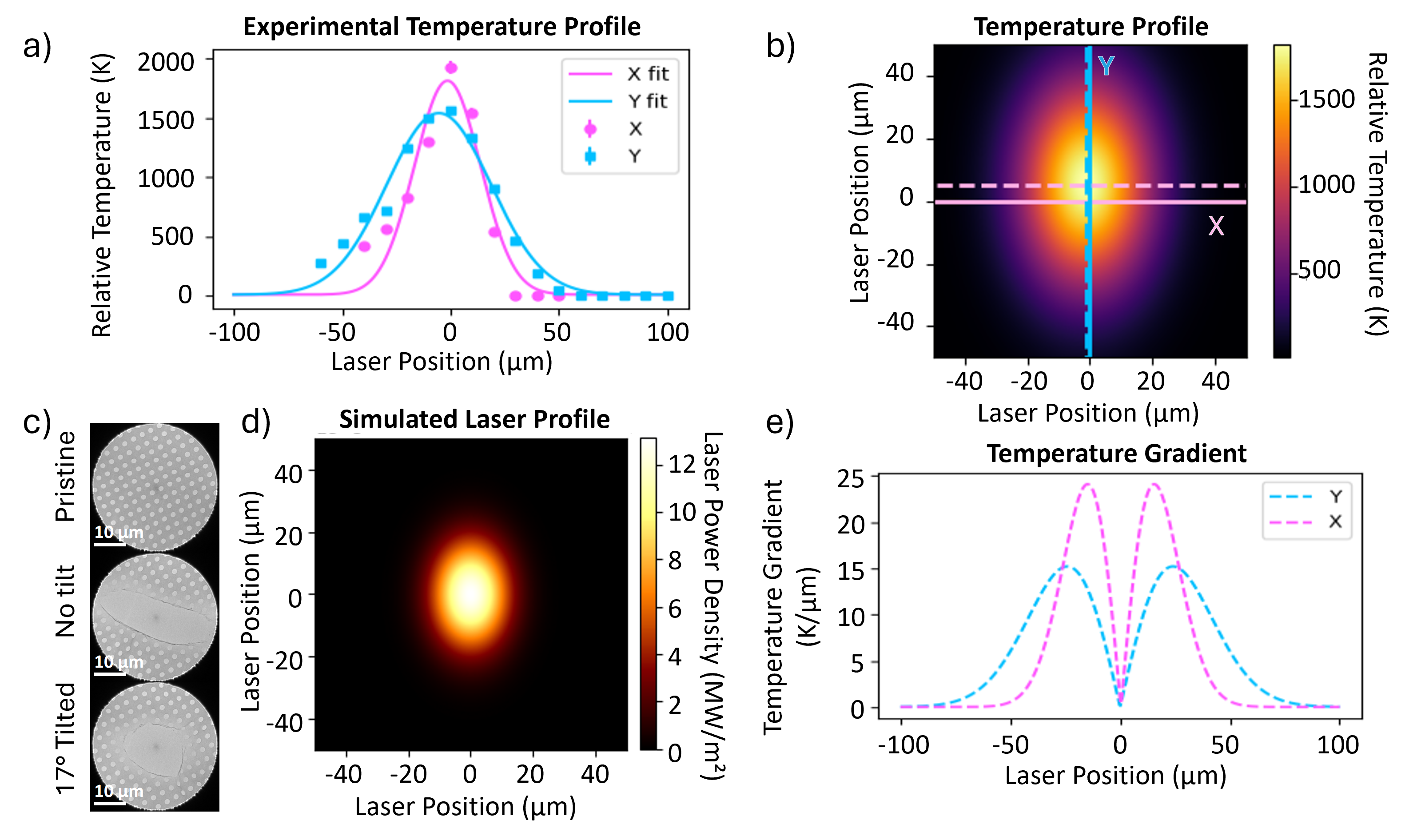} 
    \caption{Temperature and laser power distributions. a) Plot of temperature against the relative laser position along both laser motor axes (X and Y), fitted with Gaussian functions. The position axis is calibrated relative to the center of the profile as determined afterwards and shown in b. b) 2D Gaussian fit of the temperature, where solid lines indicate the measured axes in panel a and dashed lines indicate the center of the spot. The FWHM of the long axis (vertical) was 57.2 \(\mu\)m and 38.6 \(\mu\)m for the short axis (horizontal). c) Defocused ronchigram images of the Quantifoil grid before and after evaporation due to the laser, with no tilt and a 17$^\circ$ tilt angle of the sample stage. d) Profile of the laser beam power at the sample plane considering losses for the calculated spot size from simulations with a FWHM of 28.5 \(\mu\)m along the long axis (vertical) and 20.2 \(\mu\)m along the short axis (horizontal). e) Temperature gradients over the center of the laser spot (dashed lines in b) along both axes.}
    \label{fig:Temp_Grad}
\end{figure}

To determine the region of specimen heated by the laser and estimate the size of the laser spot, the laser position on the sample was scanned along the two axes of the laser motors and EELS spectra acquired for each step at a fixed sample position. The temperature calculated from PDB of the spectra at each laser position is plotted in Fig. \ref{fig:Temp_Grad}a for both laser axes and is fitted with a Gaussian function each. The resulting elliptical temperature distribution, as expected from the incidence angle of the laser, is shown in Fig. \ref{fig:Temp_Grad}b. Holes from laser evaporation of the carbon film are shown in Fig. \ref{fig:Temp_Grad}c for an untilted and 17$^\circ$ tilted sample holder demonstrating the elliptical shape due to the incidence angle. However, as the laser profile is Gaussian-shaped and not a top-hat function the size of the hole depends on the laser power and they can thus not be directly linked. The laser power profile in the sample plane, shown in Fig. \ref{fig:Temp_Grad}d, was determined by optimizing FTCS heat conduction and radiation modeling to the experimental data whilst considering laser power losses as described in Materials and Methods. The long axis of the laser spot has a full-width-at half-maximum (FWHM) of 28.5 \(\mu\)m and the short axis of 20.2 \(\mu\)m.  From this determined laser power distribution in the sample plane, the lower axis in Fig. \ref{fig:Temp_Increase}d was calculated taking into account also the power loss from the system as discussed in Materials and Methods. Two profiles of the temperature calculated gradient along the fundamental axes of the laser spot are shown in Fig. \ref{fig:Temp_Grad}e.

\begin{figure}[htbp]
    \centering
    \includegraphics[width=\linewidth]{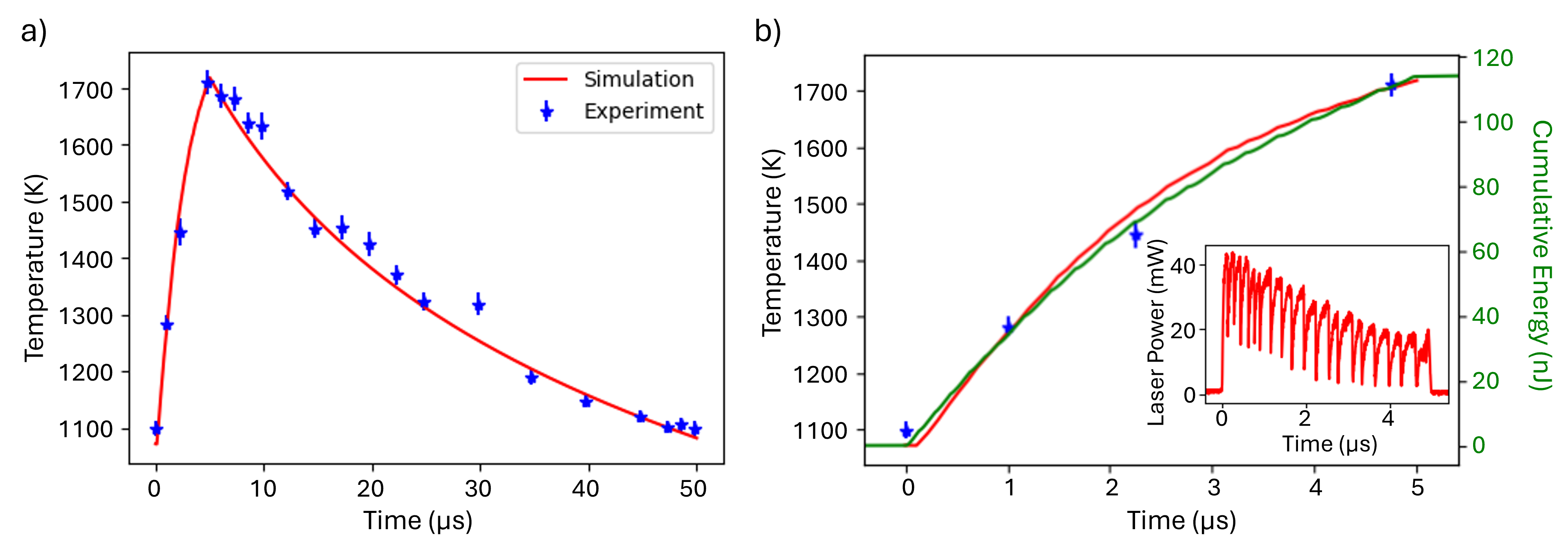} 
    \caption{Time resolved temperature measurements of amorphous carbon and comparison to simulations. a) Specimen temperature measurements (blue) during the 50 \(\mu\)s period between successive laser pulses with the simulated temperature evolution in red. b) Cumulative energy supplied by the laser pulse and specimen temperature during 5 \(\mu\)s time window of laser stimulation, where the red trace is the temperature simulation as in a. The inset shows the shape of the laser pulse scaled to the laser output power across the same time range.} 
    \label{fig:Time_Resolved_Result} 
\end{figure}

Time resolved heat transport measurements were carried out with synchronized laser excitation and detector acquisition using a signal generator, see Fig. \ref{fig:Setup}c for the setup (detailed description in Materials and Methods). An 8 nm thick amorphous carbon film was excited at a frequency of 20 kHz (50 \(\mu\)s time period) with a 5 \(\mu\)s laser pulse (10\% laser duty cycle). The center of the laser excited region was probed using the electron beam and EELS spectra collected in 500 ns wide time windows (1\% detector duty cycle) at selected time delays from the laser excitation. Spectra were collected for a total acquisition time of 200 ms (500 ns time window over 4 million cycles, 20 s total experiment time) by integration of 200 frames (20k cycles per frame) each with an exposure time of 1 ms (100 ms total experiment time) for every time delay to determine the evolution of the temperature after laser excitation. The temperature evolution of the sample is given in Fig. \ref{fig:Time_Resolved_Result}a, showing a sharp rise during the duration of the laser pulse followed by a slower decay due to thermal dissipation. Due to the low thermal conductivity of amorphous carbon, the decline in temperature is slow and a longer time window would be required to observe complete cooling to room temperature. Therefore, a steady state where the temperature oscillates between two elevated values can be observed with a \(\Delta\)T of about 610 K.
The pulse shape of the laser was measured using a silicon photodetector (see Materials and Methods) and is shown in Fig. \ref{fig:Time_Resolved_Result}b. The total energy supplied to the specimen during a laser pulse (integration of laser output power over time) was 115 nJ (determined using a power meter), but it can be seen that this energy is not homogeneously supplied to the sample. After a sharp rise at the beginning of the pulse, the average power is decaying while oscillating until the end of the pulse when it sharply drops to zero, as shown in the inset in Fig. \ref{fig:Time_Resolved_Result}b. A stronger heating from the higher laser power at the beginning of the pulse can be observed in the overlaid temperature trend of the sample during the laser pulse in Fig. \ref{fig:Time_Resolved_Result}b.

The temperature evolution of the sample across the measured time period was simulated using the FTCS heat diffusion and radiation model described in Materials and Methods. Heat conduction is the main dissipation pathway, consistent with the agreement of experiments and calculations shown by Castioni et al. \cite{Castioni2025_OrsayNL} using a pure conduction model. However, we observe a slight but significant modification and improved agreement when including radiation. The thermal conductivity and heat capacity of the amorphous carbon film were determined by optimizing these parameters in the modeling of the experimental data using gradient descent. The thermal conductivity of the film was calculated as 1.24 \(\frac{W}{m\cdot K}\) and the heat capacity determined to be 821 \(\frac{J}{kg\cdot K}\), in accordance with expectations as we will discuss below. We do not specify uncertainties for these quantities due to the lack of uncertainties in the used literature values for the modeling.

\subsection{Discussion}

The determined thermal conductivity of the amorphous carbon film of 1.24 \(\frac{W}{m\cdot K}\) is consistent with literature values varying between 0.74 and 2.2 $\frac{W}{m\cdot K}$ depending on the density and bonding within the sample \cite{Bullen2000_aC_conduct}. Likewise, the fitted heat capacity of 821 \(\frac{J}{kg\cdot K}\) is in agreement with typical values of the heat capacity of amorphous carbon ranging from 700 to 1000 \(\frac{J}{kg\cdot K}\) \cite{Hurler1995_aC_cap}. This demonstrates that the method presented here allows for reliable measurements of both the thermal conductivity and heat capacity of a specimen at the nanoscale.

In comparison to the setup by Castioni et al. \cite{Castioni2025_OrsayNL} based on a parabolic mirror in the polepiece gap, our numerical aperture of the incident laser is significantly smaller and therefore the (diffraction-limited) laser spot size at the sample is larger.
However, as we can achieve strong heating (evaporation of carbon), the power density is not a major limitation for thermal transport experiments. The magnitude of thermal gradients is generally stronger for a more focused laser spot, but we will discuss below how to enhance them for less focused laser illumination. The advantage of the setup demonstrated here is that the absence of optical elements in the polepiece gap allows for larger sample tilt angles and also thicker holders, for example \textit{in-situ} (biasing) holders whose capabilities can thus be combined with laser excitation.

Comparing the externally gated acquisition mode of the Dectris ELA detector with an event based detector such as the Timepix 3 as used by Castioni et al. \cite{Castioni2025_OrsayNL} reveals advantages and disadvantages for both. While the externally gated acquisition of the Dectris ELA enables better than 50 ns time resolution for a detector that is intrinsically not time resolved, only the electrons that hit during the gating window contribute to the signal. In an event based detector, all electrons carry temporal information which is a fundamental advantage. However, the maximum electron flux that the Timepix 3 can meaningfully interpret is quite low (80 Mhits/s for the whole chip) and therefore the beam current often needs to be reduced for the experiments and data becomes sparser. While the Timepix 3 offers a time resolution of 1.56 ns, around a factor of thirty times better than achievable here, the temporal integration required for acceptable noise levels can limit the actual time resolution. On the other hand, the Dectris ELA can quantitatively interpret signals up to 0.8 pA per pixel, which also holds for the externally gated mode and accommodates the maximum currents per pixel reached in typical highly monochromated EELS-STEM experiments. 
Therefore, the effective flux captured with the Dectris ELA can even surpass the Timepix 3 given the integration window and the available beam current if the sensitivity of the sample is not limiting. Timepix 4 supposedly allows for significantly higher fluxes, which would greatly improve this event-based detection approach.

To obtain detailed information about thermal transport in steady state or dynamically, a strong thermal gradient is needed. The measured temperature gradient induced by the laser, seen in Fig \ref{fig:Temp_Grad}e, shows a maximum of 24 K/\(\mu\)m along the short axis of the laser spot. This is approximately 7 times smaller than the average temperature gradients achieved via resistive heating (about 160 K/\(\mu\)m) by Liu et al. \cite{Liu2025_Gao_heating}. However, we propose a method to achieve much more localized heating than the size of the laser spot: depositing small absorbers at regions of interest on the sample should facilitate strong local temperature gradients. This approach works best if the sample itself is transparent at the utilized laser wavelength. The additional sample preparation can be relatively swift, uncomplicated and compatible with many sample types, for example via carbon deposition using an electron beam and a gas injection system like in a focused ion beam system. Using absorbers that are resonant with the laser excitation wavelength, such as plasmonic particles, should lead to even stronger temperature gradients.

\subsection{Conclusion}

We have developed and demonstrated an experimental framework for nanoscale thermal transport measurements in a scanning transmission electron microscope based on laser excitation and vibrational EELS thermometry. A fiber-coupled laser injection system was integrated via a modified aperture mechanism, introducing optical excitation without optical components in the polepiece gap and maintaining compatibility with a wide range of sample holders and large tilt angles.

By synchronizing pulsed laser excitation with an externally gated acquisition mode of a direct electron detector, we achieved temporal resolution about 50 ns while retaining high electron flux capability and sub-10 meV energy resolution. Local temperatures were determined using the principle of detailed balance, and thermal transport parameters were extracted by combining time-resolved thermometry with a forward-time central-space heat diffusion model including radiative losses. For amorphous carbon thin films, we obtained thermal conductivity and heat capacity values consistent with literature, validating the quantitative capability of the approach.

The presented setup enables the generation and measurement of nanoscale thermal gradients in the electron microscope without complex microfabrication. The absence of optical components in the polepiece gap facilitates integration with in-situ biasing and tomographic experiments. Future extensions using localized absorbers or wavelength-resonant nanostructures are expected to enhance spatial localization of heating and thus thermal gradients. The system can be combined with momentum-resolved vibrational EELS experiments to study the transport properties of individual phonon modes across defects and interfaces.

This instrumentation and analysis framework provides a general platform for time-resolved studies of thermal transport in nanoscale materials and devices with applications for densely integrated electronics, plasmonic devices and quantum materials.

\subsection{Acknowledgements}
The authors thank Katrina Coogan (Humboldt-Universität zu Berlin) for help with mechanical modifications of the laser injection port. This project was funded by the Deutsche Forschungsgemeinschaft (DFG, German Research Foundation) – Projektnummer 530143441 and by the Agence Nationale de la Recherche (ANR, French National Research Agency) No. ANR-23-CE42-0028-01 (project PuMMAVi).

\subsection{Conflict of interest}
J.M., O.L.K., B.P-S., A.M. and T.C.L. declare an interest in Bruker AXS LLC. L.P. and M.M. declare an interest in Dectris AG.

\bibliography{bibliography}

\end{document}